\documentclass[runningheads]{llncs}
\usepackage[T1]{fontenc}
\usepackage{graphicx}
\usepackage{amsmath}
\usepackage{csquotes}
\usepackage{listings}
\PassOptionsToPackage{hyphens}{url}
\usepackage{hyperref}
\usepackage{todonotes}
\usepackage{lineno}
\usepackage{booktabs} 

\usepackage{cleveref}       
\usepackage{subcaption}     
\usepackage{pgfplots}       
\usepackage{float}          
\usepgfplotslibrary{groupplots}
\pgfplotsset{
    questionmark/.style={
        mark=text,
        text mark={\scriptsize ?}, thick
    }
}
\pgfplotsset{compat=1.18}   

\newcommand{\marksize}{1} 

\usepackage{color}

\begin{document}
\title{Don't Blame the Model, Verify the Data: An Evaluation of SMT-based Dataset Verification  (Extended Version)}
\titlerunning{An Evaluation of SMT-based Dataset Verification}
%

\author{
Sehee Park\inst{1}  \and
Dominik Geißler\inst{1}\orcidID{0009-0008-8069-1417} \and
Andrei Aleksandrov\inst{1,2}\orcidID{0000-0002-4717-4206} \and
Kim Völlinger\inst{1}\orcidID{0000-0002-8988-0053} 
}
\authorrunning{S. Park et al.}
%
\institute{Technische Universität Berlin, Berlin, Germany\\
\email{voellinger@tu-berlin.de}
 \and
Fraunhofer FOKUS, Berlin, Germany\\
}
\maketitle              
\begin{abstract}
The EU AI Act mandates that datasets for high-risk machine learning (ML) systems meet strict quality criteria such as soundness and bias mitigation. 
While Satisfiability Modulo Theory (SMT) solving offers a formal approach to verifying these properties, 
its scalability in realistic ML settings remains unexplored.
To bridge this gap, this work presents the first large-scale empirical study of SMT-based dataset verification on two real-world ML datasets. 
We systematically evaluate how solver performance is shaped by three key dimensions: the type of data-quality property, the specification style, and the dataset encoding strategy. 
Our findings demonstrate that SMT-based verification is feasible for practical scenarios, but each dimension shapes it: 
the property type sets the tractability limit, the specification style drives scalability (exceeding $2{,}000{\times}$ for aggregate properties), 
and the encoding strategy has a systematic effect, with extracted feature columns performing best.
\end{abstract}
\section{Introduction}
\label{sec:intro}

Many failures of ML models originate directly from their training or evaluation data.
Data quality serves as a primary determinant of model reliability:
incomplete data can introduce vulnerabilities that lead to unsafe behavior in rare scenarios~\cite{schwalbe2020structuring}, 
biased data can produce discriminatory models~\cite{HANNA2025100686}, and poisoned data present severe security threats~\cite{tian2022comprehensive}. 
Consequently, data quality has become a central objective in industrial practice and AI research~\cite{data-quality-survey}.
This imperative is reinforced by emerging regulations that elevate data quality to a legal requirement.
Notably, the EU AI Act~\cite{euaiact} mandates that data used in high-risk AI systems satisfy properties 
such as completeness, representativeness, soundness, and bias mitigation.
In practice, data quality is typically assessed using libraries for surface-level validation, as task-specific semantic constraints are difficult to generalize, while
research investigates bug discovery methods using heuristics and ML models~\cite{data-quality-survey}.
However, these methods lack formal rigor. As regulatory and safety demands increase, a fundamental question arises: 
Can data quality be specified and verified using traditional formal methods?

To date, research on formal methods for ML has largely focused on post-training verification,
but recent work shifted attention to earlier stages of the ML pipeline 
to train models that are inherently more likely to satisfy specifications~\cite{intersymbolic_paper}.
Yet, the data stage remains largely unexplored,
even though training data violating a specification likely results in incorrect model behavior. 
Moreover, certain semantic properties can be ensured for the data.
This perspective is conceptually aligned with techniques like adversarial training, 
where training data are augmented by data points that promote model robustness~\cite{AugMix}.
Existing formal methods for ML data are sparse, but include
abstract interpretation-based data pre-processing~\cite{abstract1,abstract3,abstract4}, 
verified synthetic data generation~\cite{generation,10.1145/3587828.3587869}, and
most relevant to this work, a reduction of data-quality verification to SMT solving~\cite{dsverification}. 
SMT solving extends propositional satisfiability through theories such as linear integer arithmetic or arrays.
Given  a set of assertions, an SMT solver returns 
\texttt{sat} if a satisfying assignment exists, \texttt{unsat} if it does not, and \texttt{unknown} if it cannot determine the result within its resource limits.
In~\cite{dsverification}, a tabular dataset is encoded as an array $D[i][j]$, where $i$ indicates a record and $j$ a feature, 
together with a separate array for labels, and a property is expressed as a formula over these arrays. 
Verification reduces to checking whether the conjunction of the dataset encoding and the property formula is satisfiable.
However, since this approach has been evaluated 
only in a proof-of-concept setting on ten synthetic data points, 
its practicality in realistic settings remains unclear.\footnote{There has been an internal evaluation up to 120 synthetic data points, but these results are not reproducible because the dataset is not public~\cite{dsverification}.} 

To bridge this gap, we present the first large-scale empirical study of SMT-based dataset verification for real-world ML datasets. 
Using the state-of-the-art SMT solver Z3~\cite{z3}, we evaluate the approach on industrial ML benchmark datasets: German Credit (1,000 records, 24 features, binary label)~\cite{german}
and Bank Marketing (45,211 records, 16 features, binary label)~\cite{bank}.
We systematically study how solver performance is shaped by three key dimensions:
type of data-quality property, the specification style, and the dataset encoding strategy.
Our findings: 
(1) property type determines the limits of the approach with
continuous relational properties with alternating quantifiers producing irregular solving times and requiring algebraic pre-processing to extend the tractable range; 
(2) specification style profoundly impacts scalability
with gaps widening when the dataset scales and exceeding $2{,}000{\times}$ for aggregate properties; and
(3) data encoding has a systematic, style-independent effect with extracted feature columns performing best.
Appendix~\ref{appendix} includes all experiments. 
This paper makes the following contributions:
\begin{enumerate}
\item \emph{Empirical Evaluation:} scalability of SMT-based dataset verification, comparing property types, specification styles, and encoding strategies.
\item \emph{Benchmark Suite:} over 400 SMT-LIB benchmark instances for ML data quality for 22 experiments
to support future research on dataset verification.\footnote{\url{https://github.com/seheepark9954/Empirical-Evaluation-of-SMT-based-Dataset-Verification}}
\end{enumerate}

\section{Performing SMT-based Dataset Verification}
\label{sec:results}

We present the evaluation framework (Section~\ref{sec:specifying-properties}) and
the empirical evaluation (Section~\ref{sec:evaluation}).

\subsection{Evaluation Framework}
\label{sec:specifying-properties}

We introduce the three dimensions of our study.

\paragraph{Data-Quality Property Types.} 
We establish a taxonomy of data-quality properties by scope, structure, and domain, aligning conventions from databases with SMT solving.
\emph{Record-level} properties constrain each data point individually such as range validity, 
analogous to check constraints in databases~\cite{10.14778/2536258.2536262}.
These properties correspond to universal array properties~\cite{arrays-decidable}.
\emph{Dataset-level} properties constrain the whole dataset,
falling into two distinct structures.
\emph{Aggregate} properties depend on summary statistics derived from a selection of record features such as the median, comparable to database aggregation constraints~\cite{ROSS1998149}.
\emph{Relational} properties define relationships between records, similar to database denial constraints~\cite{10.14778/2536258.2536262}.
Examples include duplicate elimination and ensuring the existence of counterfactual-fair records.
Relational properties can also relate records to the domain space (e.g., domain coverage).
Underlying variables can be \emph{discrete} or \emph{continuous}, 
falling into the theories of (non-)linear integer or real arithmetic.

From this taxonomy we select four representative measures for evaluation that represent a substantial range of property types.
Table~\ref{tab:properties} maps the selected measures to four data-quality criteria of the EU AI Act (Article 10) but is not a formalization of the Article,
which states these criteria qualitatively and also mandates governance processes and contextual fit. 
SMT-based verification targets the fragment of these criteria that admits a predicate.
\begin{table}[t]
\caption{The quality measures selected for evaluation (indicated in bold), categorized by EU AI Act requirement~\cite{euaiact} and property type.}
\label{tab:properties}
\centering
\begin{tabular}{l l l}
        \toprule
\emph{EU AI Act requirement} & \emph{example measure} & \emph{example property type} \\ \midrule
soundness        & \textbf{min-max normalization}      & record-level, continuous, linear   \\ 
                 & \textbf{range validity}      & record-level, discrete, linear     \\ \hline
representativeness       & \textbf{class balance}        & aggregate, discrete, linear       \\ \hline
completeness       &   \textbf{well-distributedness}      & relational, continuous, non-linear       \\ \hline
bias mitigation       &   demographic parity     & aggregate, discrete, linear \\ \bottomrule
\end{tabular}
\end{table}
We adapt three of the four measures introduced in the foundational reduction framework~\cite{dsverification},
replacing their record-count property (a lower bound on the number of records) with range validity. Concretely, we evaluate the following data-quality measures, where throughout, $j$ denotes the fixed feature under verification and $i$ ranges over records:\footnote{
Demographic parity  serves as an example but is not evaluated separately: over a dataset it is an aggregate, discrete, linear property that reduces to per-group counting, sharing its property type with class balance.}  
\begin{enumerate}
     \item \emph{Min-max normalization} is a preprocessing transformation requiring a feature $j$ to be normalized~\cite{dsverification}:
    $\forall i.\; (0 \le i < m) \;\to\; -1 \le D[i][j] < 1$. 
    \item \emph{Range validity} ensures that an unprocessed feature falls within a predefined interval~\cite{data-quality-survey}: $\forall i.\; (0 \le i < m) \;\to\; low \le D[i][j] < high$. Although logically similar to min-max normalization, in our experiments, we apply different dataset encoding strategies to them. 
   \item \emph{Class balance} ensures that no class label is underrepresented~\cite{dsverification}.
Each label must appear at least $\frac{m}{l \cdot B}$ times, where $m$ is the number of records, $l$ is the number of distinct labels and $B$ is a tolerance parameter. The tolerance $B$ is a user-supplied parameter that defines the acceptable degree of class imbalance for a given intended purpose. Its value affects the satisfying assignment but not the structure of the encoding or its solving cost.
The property requires an aggregation (counting) over all records.
   \item \emph{Well-distributedness} dictates that records cover the feature space without large gaps. 
   Since solving immediately resulted in \texttt{unknown} in the original work~\cite{dsverification},
    we project it to a one-dimensional case, a necessary condition for multi-dimensional verification.
    For feature $j$, no point $p$ exists who is more distant to every record than threshold $d$:
$\neg \exists p \;.\; \forall i.\; (0 \le i < m)  \;\to\; 
\mathrm{dist}(p, D[i][j]) > d$ with $\mathrm{dist}$ being the Euclidean distance. 
\end{enumerate}
These four measures were selected as representatives of the taxonomy's property types, and our results transfer along those types.

\paragraph{Specification Styles.}
Each property can be expressed in multiple logically equivalent ways in SMT-LIB 2.
We evaluate three distinct specification styles, while the applicable styles depend on the property:
\begin{enumerate}
     \item The \emph{baseline} style is the original quantified formulation from the reduction framework~\cite{dsverification}. While highly readable, it ignores solver efficiency. 
     \item The \emph{recursive} style constructs a recursive function which iterates over the dataset. For class balance, the baseline style already employs recursion~\cite{dsverification}.
     \item The \emph{grounding} style exploits that a dataset is finite. It completely eliminates quantifiers and recursion by expanding the conditions into explicit conjunctions over the index range, placing the problem into a decidable fragment.
\end{enumerate}     
Note that because well-distributedness quantifies existentially over a continuous feature space, it cannot be expressed using the grounding style. 
Furthermore, non-linear real arithmetic is decidable only with doubly exponential worst-case complexity via cylindrical algebraic decomposition (CAD)~\cite{cad-doubly}.
Since CAD expects polynomial input, roots require algebraic transformations,
automatically performed during preprocessing~\cite{z3internals}.
We apply a targeted \emph{optimization technique} that 
squares both sides of the distance threshold equation. 
This optimization eliminates non-polynomial behavior
to bypass volatile solver preprocessing heuristics, which are at work to avoid worst-case complexity.
Additionally, it reduces the problem to a quadratic case, 
which often has solutions more efficient than CAD and frequently quantifier elimination can be accelerated~\cite{root-elimination}.

\paragraph{Datasets Encodings.}
We evaluate three tabular-data encoding strategies that are alternative SMT-LIB representations of the same dataset:
\begin{enumerate}
     \item \emph{Nested Array:} The original encoding~\cite{dsverification} employs a two-dimensional array of type \texttt{Array Int (Array Int \_)}, 
     where the outer array indexes a record and the inner array indexes its features. 
     Quantification over a one-dimensional array is known to be challenging~\cite{reynolds2013}, and nesting requires an additional layer of resolution of array axioms~\cite{z3internals}. 
     \item \emph{Column-Slice:} This encoding represents only the feature column relevant to a given property (in the SMT model), as an array of type \texttt{Array Int \_}.
     \item \emph{Nested Column-Slice:} A nested array that wraps only the relevant feature column, serving as a control to isolate nesting cost from multi-column formula-size effects. 
\end{enumerate} 

\subsection{Empirical Evaluation}
\label{sec:evaluation}

We investigate one research question for each evaluation dimension:
\begin{itemize}
    \item[]\textbf{RQ1}: Which data-quality property types remain tractable for SMT-based verification, and to what extent can algebraic preprocessing extend tractability for the challenging case: non-linear continuous relational properties?
    \item[]\textbf{RQ2}: How does the specification style affect solver performance and scalability, and how do these effects evolve with dataset size?
    \item[]\textbf{RQ3}: How does dataset encoding influence solver performance?
\end{itemize}
\begin{table}[t]
    \centering
    \caption{Categorization of the 22 experiments with references to their results.}
    \label{tab:experiments}
    \begin{tabular}{lll}
        \toprule
           \textbf{Property}      & \textbf{German Credit Benchmark} & \textbf{Bank Marketing Benchmark} \\
        \midrule
        \textbf{Min-max normal.} & baseline, recursive, grounding  & baseline, recursive, grounding  \\
       (specification style)  & as nested column data (Fig.~\ref{ap:min-max-german})  & as nested column data (Fig.~\ref{fig:local-properties})  \\\hline
        
        \textbf{Range validity} & baseline, recursive, grounding  &  baseline, recursive, grounding  \\ 
      (specification style)  & as nested array data (Fig.~\ref{ap:range-german})  &  as nested array data (Fig.~\ref{fig:global-properties}) \\\hline
        
        \textbf{Class balance} & recursive baseline, grounding  &  recursive baseline, grounding \\ 
         (specification style) & as column data (Fig.~\ref{ap:class-german})  &  as column data (Fig.~\ref{fig:local-properties}) \\\hline
        
        \textbf{Well-distributed} &  & baseline, optimized baseline \\
         (optimization and &   & as nested array data and \\ 
        encoding strategy) &  & as column data (Fig.~\ref{fig:global-properties} and~\ref{ap:well-distributedness}))\\ \hline
        
        \textbf{Range validity} &  &  nested array, column and  \\
         (encoding strategy)                       &  &  nested column data\\
                               &  &  for grounding (Fig.~\ref{fig:global-properties} and~\ref{ap:dataset-encodings})\\
        \bottomrule
    \end{tabular}
\end{table}

\paragraph{Experimental Set-up.} 
We conducted 22 experiments (see \Cref{tab:experiments}), progressively increasing the number of records ($m$) to evaluate scalability. 
For min-max normalization and range validity, we compared all specification styles. For class balance, we compared the recursive baseline and grounding style on both benchmarks.
Well-distributedness was evaluated exclusively on the larger Bank Marketing benchmark, because quantified non-linear terms dominate scalability limits and make it the most demanding property. 
Although min-max normalization and range validity share a similar logical structure, they were studied using different data encodings. To isolate encoding effects, we fixed the property (range validity), specification style (grounding), and dataset (Bank Marketing), varying only the encoding strategy.
Before performance evaluation, we validated that the specification styles agree on the verification verdict on each property and dataset instance.
Experiments were run on a MacBook Air (Apple M2, 8 GB RAM) with Z3 4.15.3 with default settings. 
Reported runtimes are arithmetic means of three runs. CPU usage remained above 95\% to avoid measurement artifacts, and the timeout was set to 10 minutes.

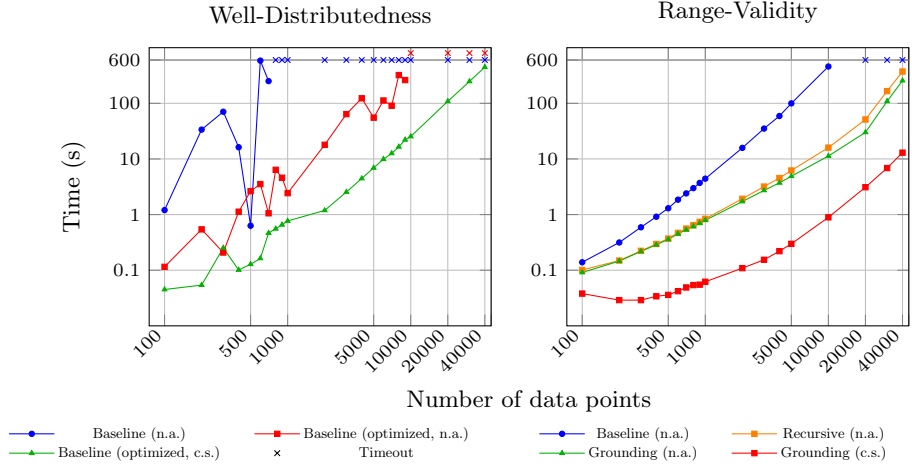
\begin{figure}
    \centering
    \begin{tikzpicture}
    \begin{groupplot}[
        group style={
        group size=2 by 1,
        horizontal sep=1cm,
    },
    width=0.5\textwidth,
    xmode=log,
    ymode=log,
    xmin=75,
    xmax=45000,
    ymin=0.01,
    ymax=1000,
                   xlabel={},
                ylabel={},
                label style={font=\small},
                tick label style={font=\scriptsize},
                grid=major,
                legend pos=south east,
                legend style={nodes={scale=0.6, transform shape}},
                log ticks with fixed point,
                xtick={100,500,1000,5000,10000,20000,40000,45000},
                xticklabels={100,500,1000,5000,10000,20000,40000},
                    xticklabel style={
                rotate=45,
               anchor=east,
               },
               ytick={0.1,1,10,100, 600},
                yticklabels={0.1,1,10,100, 600},
             legend columns=2,
            legend style={draw=none, /tikz/every even column/.append style={column sep=0.4cm},
            yshift=-20mm,
            },
    ]

\nextgroupplot[title={{Well-Distributedness}}]

           \addplot[color=blue, mark=*, mark size=\marksize pt] coordinates {
                (100,1.204) (200,33.720) (300,70.563) (400,16.245) 
                (500,0.631) (600,585.890) (700,250.567)
            };
            \addlegendentry{Baseline (n.a.)}

            \addplot[color=red, mark=square*, mark size=\marksize pt] coordinates {
                (100,0.115) (200,0.542) (300,0.208) (400,1.126) 
                (500,2.637) (600,3.529) (700,1.055) (800,6.381) 
                (900,4.596) (1000,2.427) (2000,17.968) (3000,64.020) 
                (4000,123.153) (5000,55.265) (6000,112.407) (7000,90.590) 
                (8000,322.107) (9000,261.887) 
            };
            \addlegendentry{Baseline (optimized, n.a.)}


    \addplot[color=green!70!black, mark=triangle*, mark size=\marksize pt] coordinates {
        (100,0.045)
        (200,0.054)
        (300,0.252)
        (400,0.101)
        (500,0.128)
        (600,0.163)
        (700,0.466)
        (800,0.555)
        (900,0.655)
        (1000,0.768)
        (2000,1.189)
        (3000,2.545)
        (4000,4.451)
        (5000,6.917)
        (6000,9.897)
        (7000,12.590)
        (8000,16.525)
        (9000,22.064)
        (10000,25.341)
        (20000,108.587)
        (30000,247.307)
        (40000,448.157)
    };
    \addlegendentry{Baseline (optimized, c.s.)}


    \addlegendimage{black,mark=x,,only marks,mark size=\marksize * 1.5 pt}
\addlegendentry{Timeout}
             \addplot[
color=blue,
    mark=x,
    only marks, mark size=\marksize*1.5 pt
] coordinates {
    (800,600) (900,600) (1000, 600)
(2000, 600)
(3000, 600)
(4000, 600)
(5000, 600)
(6000, 600)
(7000, 600)
(8000, 600)
(9000, 600)
(10000, 600)
(20000, 600)
(30000, 600)
(40000, 600)
};
             \addplot[
color=red,
    mark=x,
    only marks, mark size=\marksize*1.5 pt
] coordinates {
(10000,800)
(20000,800) (30000,800) (40000,800)
};

\addplot[gray, domain=\pgfkeysvalueof{/pgfplots/xmin}:
                               \pgfkeysvalueof{/pgfplots/xmax}]
    {600};

\nextgroupplot[title={{Range-Validity}}]
            \addplot[color=blue, mark=*, mark size=\marksize pt] coordinates {
                (100,0.139) (200,0.315) (300,0.591) (400,0.916) (500,1.300)
                (600,1.854) (700,2.402) (800,2.985) (900,3.707) (1000,4.422)
                (2000,15.829) (3000,35.152) (4000,59.117) (5000,99.707) (10000,458.763)
            };
            \addlegendentry{Baseline (n.a.)}
            \addplot[color=orange, mark=square*, mark size=\marksize pt] coordinates {
                (100,0.101) (200,0.149) (300,0.223) (400,0.296) (500,0.370)
                (600,0.469) (700,0.560) (800,0.644) (900,0.742) (1000,0.839)
                (2000,1.923) (3000,3.187) (4000,4.534) (5000,6.203) (10000,15.985)
                (20000,51.174) (30000,165.577) (40000,372.547)
            };
            \addlegendentry{Recursive (n.a.)}
            \addplot[color=green!70!black, mark=triangle*, mark size=\marksize pt] coordinates {
                (100,0.091) (200,0.146) (300,0.218) (400,0.287) (500,0.359)
                (600,0.451) (700,0.533) (800,0.611) (900,0.696) (1000,0.791)
                (2000,1.718) (3000,2.721) (4000,3.718) (5000,4.894) (10000,11.280)
                (20000,30.040) (30000,108.980) (40000,256.190)
            };
            \addlegendentry{Grounding (n.a.)}
    \addplot[color=red, mark=square*, mark size=\marksize pt] coordinates {
        (100,0.038)
        (200,0.029)
        (300,0.029)
        (400,0.034)
        (500,0.036)
        (600,0.042)
        (700,0.049)
        (800,0.054)
        (900,0.055)
        (1000,0.062)
        (2000,0.109)
        (3000,0.154)
        (4000,0.220)
        (5000,0.298)
        (10000,0.892)
        (20000,3.105)
        (30000,6.866)
        (40000,12.972)
    };
    \addlegendentry{Grounding (c.s.)}

                 \addplot[
color=blue,
    mark=x,
    only marks, mark size=\marksize*1.5 pt
] coordinates {
    (20000,600) (30000,600) (40000,600)
};

\addplot[gray, domain=\pgfkeysvalueof{/pgfplots/xmin}:
                               \pgfkeysvalueof{/pgfplots/xmax}]
    {600};

    \end{groupplot}

\node[rotate=90] at
    ($(group c1r1.south west)!0.5!(group c1r1.north west)+(-1cm,0)$)
    {Time (s)};
\node at ($(group c1r1.south)!0.5!(group c2r1.south)+(0,-1cm)$)
    {Number of data points};
\end{tikzpicture}
    \caption{Runtime for well-distributedness (left) and range-validity (right) on the Bank Marketing data, displayed in log-log scale. In experiments labeled by n.a.\ (c.s.) the data is represented by a nested array (column slice). The tabular data of these experiments are available in \Cref{ap:well-distributedness} and \Cref{ap:table-range-validity}.}\label{fig:global-properties}
\end{figure}

\paragraph{Results on RQ1.}
Among the four property types, only the continuous relational property, well-distributedness, falls outside the grounding style because it quantifies over a continuous feature space. It therefore defines the tractability boundary, whereas record-level and aggregate properties remain tractable at real-world scale under grounding (RQ2).
Unlike the other properties, well-distributedness exhibits highly non-monotonic solving times.
With the baseline formulation and nested-array enconding, runtime increases from 0.63 seconds at $m = 500$, to 585.9 seconds at $m = 600$, to 250.6 seconds at $m = 700$. 
This volatility likely stems from the interaction of Z3’s non-linear arithmetic procedures~\cite{nla-heuristics,z3internals}, CAD-related heuristics, quantifier alternation~\cite{ematching}, and the known difficulty of quantified reasoning over (nested) arrays~\cite{reynolds2013}.
Our root-elimination optimization transforms the problem into a quadratic polynomial form,
educing reliance on preprocessing heuristics~\cite{z3internals} and
improving non-linear arithmetic reasoning and quantifier instantiation~\cite{root-elimination}.
Using the same nested-array encoding, it extends the tractable range from 700 to 10,000 records, although runtime irregularities remain.

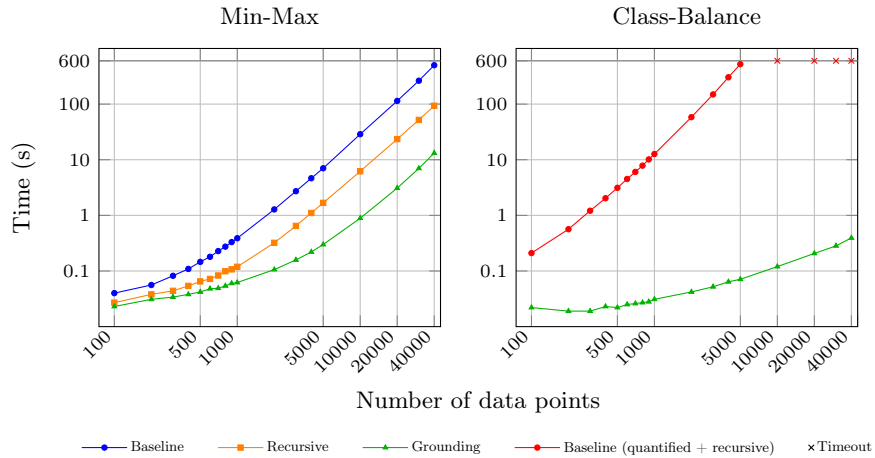
\begin{figure}
    \centering
    \begin{tikzpicture}
    \begin{groupplot}[
        group style={
        group size=2 by 1,
        horizontal sep=1cm,
    },
    width=0.5\textwidth,
    xmode=log,
    ymode=log,
    xmin=75,
    xmax=45000,
    ymin=0.01,
    ymax=1000,
                   xlabel={},
                ylabel={},
                label style={font=\small},
                tick label style={font=\scriptsize},
                grid=major,
                legend pos=south east,
                legend style={nodes={scale=0.6, transform shape}},
                log ticks with fixed point,
                xtick={100,500,1000,5000,10000,20000,40000,45000},
                xticklabels={100,500,1000,5000,10000,20000,40000},
                    xticklabel style={
                rotate=45,
               anchor=east,
               },
               ytick={0.1,1,10,100, 600},
                yticklabels={0.1,1,10,100, 600},
             legend to name=grouplegend1,
             legend columns=-1,
            legend style={draw=none, /tikz/every even column/.append style={column sep=0.4cm}},
    ]

    \nextgroupplot[title={{Min-Max}}]
                 \addplot[color=blue, mark=*, mark size=\marksize pt] coordinates {
                (100,0.040) (200,0.056) (300,0.082) (400,0.109)
                (500,0.146) (600,0.180) (700,0.228) (800,0.274)
                (900,0.331) (1000,0.388) (2000,1.279) (3000,2.718)
                (4000,4.648) (5000,7.031) (10000,28.784)
                (20000,114.133) (30000,263.310) (40000,500.127)
            };
            \addlegendentry{Baseline}

            \addplot[color=orange, mark=square*, mark size=\marksize pt] coordinates {
                (100,0.027) (200,0.038) (300,0.044) (400,0.054)
                (500,0.065) (600,0.072) (700,0.083) (800,0.099)
                (900,0.107) (1000,0.119) (2000,0.321) (3000,0.646)
                (4000,1.105) (5000,1.677) (10000,6.205)
                (20000,23.499) (30000,51.932) (40000,93.410)
            };
            \addlegendentry{Recursive}

            \addplot[color=green!70!black, mark=triangle*, mark size=\marksize pt] coordinates {
                (100,0.023) (200,0.031) (300,0.034) (400,0.038)
                (500,0.042) (600,0.048) (700,0.049) (800,0.054)
                (900,0.060) (1000,0.062) (2000,0.106) (3000,0.158)
                (4000,0.220) (5000,0.299) (10000,0.888)
                (20000,3.093) (30000,6.994) (40000,13.160)
            };
\addplot[gray, domain=\pgfkeysvalueof{/pgfplots/xmin}:
                               \pgfkeysvalueof{/pgfplots/xmax}]
    {600};

    \nextgroupplot[title={{Class-Balance}}]
            \addlegendimage{blue,mark=*,mark size=\marksize pt}
\addlegendentry{Baseline}
            \addlegendimage{orange,mark=square*,mark size=\marksize pt}
            \addlegendentry{Recursive}
            \addlegendimage{color=green!70!black, mark=triangle*, mark size=\marksize pt}
             \addlegendentry{Grounding}

             \addlegendimage{color=red, mark=*, mark size=\marksize pt]}
             \addlegendentry{Baseline (quantified + recursive)}
    \addlegendimage{black,mark=x,,only marks,mark size=\marksize * 1.5 pt}
\addlegendentry{Timeout}

            \addplot[color=red, mark=*, mark size=\marksize pt] coordinates {
                (100,0.210) (200,0.564) (300,1.212) (400,2.033) (500,3.128)
                (600,4.511) (700,6.047) (800,7.835) (900,10.133) (1000,12.671)
                (2000,58.257) (3000,149.227) (4000,303.570) (5000,521.943)
            };
            \addplot[color=green!70!black, mark=triangle*, mark size=\marksize pt] coordinates {
                (100,0.022) (200,0.019) (300,0.019) (400,0.023) (500,0.022)
                (600,0.025) (700,0.026) (800,0.027) (900,0.028) (1000,0.031)
                (2000,0.042) (3000,0.052) (4000,0.064) (5000,0.071) (10000,0.120)
                (20000,0.207) (30000,0.283) (40000,0.393)
            };

 \addplot[
color=red,
    mark=x,
    only marks, mark size=\marksize*1.5 pt
] coordinates {
(10000, 600)
(20000, 600)
(30000, 600)
(40000, 600)
};
\addplot[gray, domain=\pgfkeysvalueof{/pgfplots/xmin}:
                               \pgfkeysvalueof{/pgfplots/xmax}]
    {600};
    \end{groupplot}

\node[rotate=90] at
    ($(group c1r1.south west)!0.5!(group c1r1.north west)+(-1cm,0)$)
    {Time (s)};
\node at ($(group c1r1.south)!0.5!(group c2r1.south)+(0,-1cm)$)
    {Number of data points};

\node at ($(group c1r1.south)!0.5!(group c2r1.south)+(0,-1.6cm)$)
    {\pgfplotslegendfromname{grouplegend1}};
\end{tikzpicture}
    \caption{Runtime for min-max normalization (left) and class balance (right) on the Bank Marketing data, displayed in log-log scale. The tabular data of these experiments are available in \Cref{ap:table-min-max} and \Cref{ap:table-class-balance}.}
    \label{fig:local-properties}
\end{figure}

\paragraph{Results on RQ2.} 
The results confirm expected performance differences between specification styles: across all properties and benchmarks, grounding style performs best, followed by the recursive style, while the baseline style performs worst.
For record-level properties, the gap between baseline and grounding reaches 39x at $m=30,000$ for min-max normalization on the Bank Marketing data.
The effect is even stronger for the aggregate property class balance.
On the German Credit data at $m = 1000$, 
the recursive baseline takes 103.4 seconds, 
while the grounding style requires just 0.049 seconds, a 2,000x speedup. 
On the Bank Marketing data, the baseline style times out at $m \geq 10,000$, 
while the grounding style completes in under 0.4 seconds even at $m = 40,000$.
This divergence arises because the baseline relies on recursive counting and outer quantification over labels,
forcing the solver to alternate between recursive unfolding and heuristic E-matching~\cite{z3internals,ematching}.
In contrast, the grounding style creates a decidable problem.

\paragraph{Results on RQ3.} 
Switching the data encoding from nested arrays to a column encoding for well-distributedness
substantially stabilizes solver behavior and extends tractability from 10,000 records to 40,000 records.
Even the unoptimized baseline benefits considerably, 
although some irregularities persist.\footnote{The graph (baseline and column) can be found in Figure~\ref{ap:well-distributedness} of the appendix.}  
Cross-property comparisons show that range validity (nested-array encoding) is consistently slower than min-max normalization (column-slice encoding).
In the nested-array representation, feature access requires traversing a multi-column structure and handling an additional layer of array axioms. This overhead grows with the number of dataset features.
Consequently, the effect is stronger for German Credit (24 features) than for the
Bank Marketing (16 features).
For example, on German Credit at $m=1000$ (baseline), 
solving for min-max normalization takes 0.331 seconds and
for range validity 7.78 seconds, a 23x slowdown.
The slowdown appears across all specification styles and is largest for grounding, followed by baseline and recursive styles.
At $m=800$ and $m=900$ on German Credit, 
range validity is 31x slower than min-max normalization under grounding,23x under the baseline, and 16x under the recursive style.
We currently lack an explanation for the difference between specification styles.
To isolate encoding effects, we fixed the property to range validity and the specification style to grounding. The results confirm a substantial gap between nested-array and (nested) column-slice encodings, with slowdown factors similar to those observed in the grounding-style comparison above.
Finally, the control experiment shows that nesting itself has little impact. At $m=40,000$ solving time is 13.307 seconds for nested column-slice data and  12.972 seconds for column-slice data, a difference of 2.6\%.\footnote{The control graph (nested-column) can be found in Figure~\ref{ap:dataset-encodings} in the appendix.}
The primary overhead therefore arises from the multi-column access pattern rather than the nesting structure.
\section{Conclusion and Future Work}
\label{s:conclusion}

As practical guidance, express record-level and aggregate properties in the grounding style, 
algebraically preprocess continuous relational properties, and encode the relevant feature as a column slice. 
However, our results are limited by running a single solver on a single consumer machine, which is why the reported times are best read as relative comparisons.
Several future directions arise from the findings:
evaluation of a portfolio of SMT solvers;
a domain-specific language compiling to grounding-style SMT-LIB with automatic root elimination,
and a primitive operation for column projection;
incremental verification for growing datasets via Z3's push/pop interface;
counterexample-guided data repair for \texttt{unsat} models;
parallel verification for record-level properties;
exploration of cross-dataset properties such as train/test data leakage;
and extension to time-series and graph-structured data.


%
%
%
\bibliographystyle{splncs04}
\bibliography{bibliography}
\appendix
\section{Experimental Evaluation}
\label{appendix}
\begin{figure}
\centering
\begin{minipage}[t]{0.35\textwidth}
\vspace{0pt}
    \centering
            \begin{tabular}{cccc}
        \toprule
        $m$ & Baseline & Recursive &Grounding  \\
        \midrule
        100 & 0.048 & 0.025 & 0.021  \\
        200 & 0.047 & 0.031 & 0.027  \\
        300 & 0.069 & 0.039 & 0.029  \\
        400 & 0.091 & 0.049 & 0.033  \\
        500 & 0.116 & 0.058 & 0.036  \\
        600 & 0.153 & 0.071 & 0.041  \\
        700 & 0.188 & 0.083 & 0.044  \\
        800 & 0.229 & 0.090 & 0.047  \\
        900 & 0.278 & 0.107 & 0.053  \\
        1000 & 0.326 & 0.117 & 0.060 \\
        \bottomrule
        \end{tabular}
\end{minipage}
\hfill
\begin{minipage}[t]{0.50\textwidth}
\vspace{0pt}
    \centering
    \begin{tikzpicture}
                \begin{axis}[
    width=\linewidth,
    xmin=75,
    xmax=1010,
    ymin=0,
    ymax=0.5,
                   xlabel={Number of data points},
                ylabel={Time (s)},
                label style={font=\small},
                tick label style={font=\scriptsize},
                grid=major,
                legend pos=south east,
                legend style={nodes={scale=0.6, transform shape}},
                log ticks with fixed point,
                xtick={100,200,300,400,500,600,700,800,900,1000},
        xticklabels={100,200,300,400,500,600,700,800,900,1000},
                    xticklabel style={
                rotate=45,
               anchor=east,
               },
               ytick={0.1, 0.2, 0.3, 0.4, 0.5},
                yticklabels={0.1, 0.2, 0.3, 0.4, 0.5},
             legend columns=2,
            legend style={draw=none, /tikz/every even column/.append style={column sep=0.4cm},
            yshift=-20mm,
            },
    ]

            \addplot[color=blue, mark=*, mark size=\marksize pt] coordinates {
                (100,0.048) (200,0.047) (300,0.069) (400,0.091) (500,0.116)
                (600,0.153) (700,0.188) (800,0.229) (900,0.278) (1000,0.326)
            };
            \addlegendentry{Baseline}

            \addplot[color=orange, mark=square*, mark size=\marksize pt] coordinates {
                (100,0.025) (200,0.031) (300,0.039) (400,0.049) (500,0.058)
                (600,0.071) (700,0.083) (800,0.090) (900,0.107) (1000,0.117)
            };
            \addlegendentry{Recursive}

            \addplot[color=green!70!black, mark=triangle*, mark size=\marksize pt] coordinates {
                (100,0.021) (200,0.027) (300,0.029) (400,0.033) (500,0.036)
                (600,0.041) (700,0.044) (800,0.047) (900,0.053) (1000,0.060)
            };
            \addlegendentry{Grounding}

            \end{axis}
        \end{tikzpicture}
\end{minipage}
\caption{Min-Max normalization on German Credit data. Dataset encoding is a nested column-slice.}
\label{ap:min-max-german}
\end{figure}
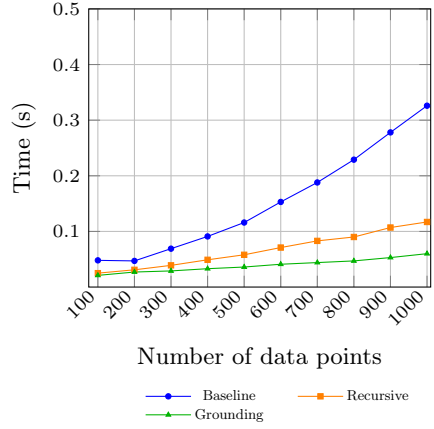

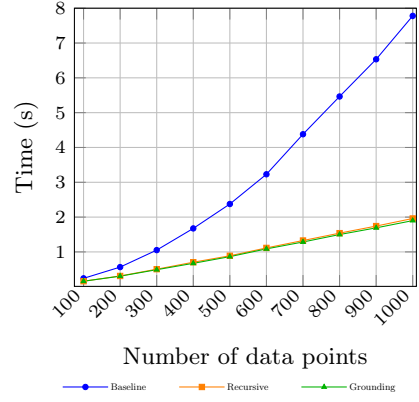
\begin{figure}
\centering
\begin{minipage}[t]{0.35\textwidth}
\vspace{0pt}
    \centering
    \begin{tabular}{cccc}
        \toprule
        $m$ & Baseline & Recursive & Grounding \\
        \midrule
        100 & 0.241 & 0.158 & 0.158  \\
        200 & 0.564 & 0.309 & 0.303  \\
        300 & 1.052 & 0.504 & 0.489  \\
        400 & 1.675 & 0.704 & 0.672  \\
        500 & 2.376 & 0.889 & 0.864  \\
        600 & 3.230 & 1.115 & 1.087  \\
        700 & 4.380 & 1.327 & 1.285  \\
        800 & 5.464 & 1.539 & 1.499  \\
        900 & 6.533 & 1.742 & 1.691  \\
        1000 & 7.780 & 1.960 & 1.899 \\
        \bottomrule
\end{tabular}
\end{minipage}
\hfill
\begin{minipage}[t]{0.50\textwidth}
\vspace{0pt}
    \centering
           \begin{tikzpicture}
            \begin{axis}[
    width=\linewidth,
    xmin=75,
    xmax=1010,
    ymin=0.01,
    ymax=8,
                   xlabel={Number of data points},
                ylabel={Time (s)},
                label style={font=\small},
                tick label style={font=\scriptsize},
                grid=major,
                legend pos=south east,
                legend style={nodes={scale=0.4, transform shape}},
                legend style={draw=none, /tikz/every even column/.append style={column sep=0.4cm},
            yshift=-16mm,
            },
                log ticks with fixed point,
                xtick={100,200,300,400,500,600,700,800,900,1000},
                xticklabels={100,200,300,400,500,600,700,800,900,1000},
                    xticklabel style={
                rotate=45,
               anchor=east,
               },
               ytick={1,2,3,4,5,6,7,8},
                yticklabels={1,2,3,4,5,6,7,8},
             legend columns=-1,
    ]
            \addplot[color=blue, mark=*, mark size =\marksize pt] coordinates {
                (100,0.241) (200,0.564) (300,1.052) (400,1.675) (500,2.376)
                (600,3.230) (700,4.380) (800,5.464) (900,6.533) (1000,7.780)
            };
            \addlegendentry{Baseline}
            \addplot[color=orange, mark=square*, mark size =\marksize pt] coordinates {
                (100,0.158) (200,0.309) (300,0.504) (400,0.704) (500,0.889)
                (600,1.115) (700,1.327) (800,1.539) (900,1.742) (1000,1.960)
            };
            \addlegendentry{Recursive}
            \addplot[color=green!70!black, mark=triangle*, mark size =\marksize pt] coordinates {
                (100,0.158) (200,0.303) (300,0.489) (400,0.672) (500,0.864)
                (600,1.087) (700,1.285) (800,1.499) (900,1.691) (1000,1.899)
            };
            \addlegendentry{Grounding}

            \end{axis}
        \end{tikzpicture}
\end{minipage}
\caption{Range validity on the German Credit data. Dataset encoding is a nested array.}
\label{ap:range-german}
\end{figure}

\begin{figure}
\centering
\begin{minipage}[t]{0.35\textwidth}
\vspace{0pt}
    \centering
     \begin{tabular}{cccc}
        \toprule
        $m$ & Baseline & Grounding \\
           \midrule
        100 & 0.395 & 0.024  \\
        200 & 1.874 & 0.023  \\
        300 & 4.043 & 0.026  \\
        400 & 11.383 & 0.029 \\
        500 & 17.006 & 0.032 \\
        600 & 26.166 & 0.035 \\
        700 & 41.206 & 0.039 \\
        800 & 60.823 & 0.042 \\
        900 & 67.703 & 0.045 \\
        1000 & 103.443 & 0.049  \\
        \bottomrule
\end{tabular}
\end{minipage}
\hfill
\begin{minipage}[t]{0.50\textwidth}
\vspace{0pt}
    \centering
    \begin{tikzpicture}
            \begin{axis}[
    width=\linewidth,
    ymode=log,
    xmin=75,
    xmax=1010,
    ymin=0.01,
    ymax=1000,
                   xlabel={Number of data points},
                ylabel={Time (s)},
                label style={font=\small},
                tick label style={font=\scriptsize},
                grid=major,
                legend pos=south east,
                legend style={nodes={scale=0.6, transform shape}},
                log ticks with fixed point,
                xtick={100,200,300,400,500,600,700,800,900,1000},
                xticklabels={100,200,300,400,500,600,700,800,900,1000},
                    xticklabel style={
                rotate=45,
               anchor=east,
               },
               ytick={0.1,1,10,100, 600},
                yticklabels={0.1,1,10,100, 600},
             legend columns=2,
            legend style={draw=none, /tikz/every even column/.append style={column sep=0.4cm},
            yshift=-16mm,
            },
    ]
            \addplot[color=blue, mark=*,  mark size=\marksize pt] coordinates {
                (100,0.395) (200,1.874) (300,4.043) (400,11.383) (500,17.006)
                (600,26.166) (700,41.206) (800,60.823) (900,67.703) (1000,103.443)
            };
            \addlegendentry{Baseline}
            \addplot[color=green!70!black, mark=triangle*, mark size=\marksize pt] coordinates {
                (100,0.024) (200,0.023) (300,0.026) (400,0.029) (500,0.032)
                (600,0.035) (700,0.039) (800,0.042) (900,0.045) (1000,0.049)
            };
            \addlegendentry{Grounding}

            \addplot[gray, domain=\pgfkeysvalueof{/pgfplots/xmin}:
                               \pgfkeysvalueof{/pgfplots/xmax}]
    {600};
            \end{axis}
        \end{tikzpicture}
\end{minipage}
\caption{Class balance on the German Credit data. Dataset encoding is a clomun-slice. The plot is displayed in log-scale.}
\label{ap:class-german}
\end{figure}
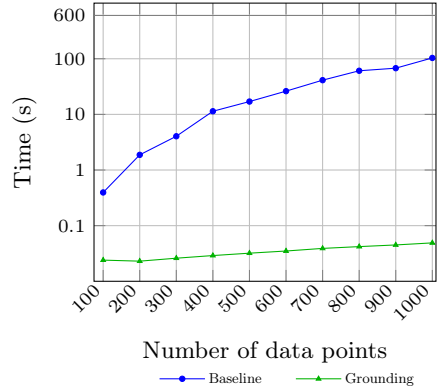

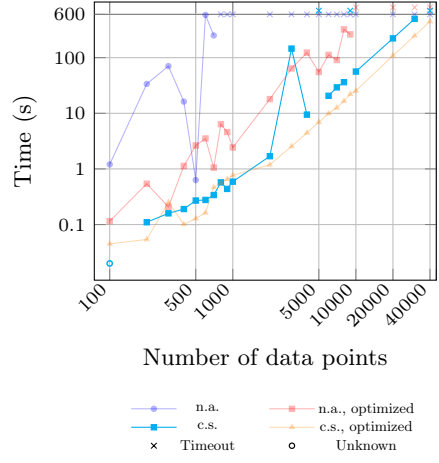
\begin{figure}
\centering
\begin{minipage}[t]{0.35\textwidth}
\vspace{0pt}
    \centering
     \begin{tabular}{ccccc}
     \toprule
        m & n.a. & n.a. (opt) & c.s. & c.s. (opt) \\     \midrule
        100 & 1.204 & 0.115 & \texttt{UK} & 0.045 \\ 
        200 & 33.720 & 0.542 & 0.110 & 0.054 \\ 
        300 & 70.563 & 0.208 & 0.159 & 0.252 \\ 
        400 & 16.245 & 1.126 & 0.190 & 0.101 \\ 
        500 & 0.631 & 2.637 & 0.270 & 0.128 \\ 
        600 & 585.890 & 3.529 & 0.277 & 0.163 \\ 
        700 & 250.567 & 1.055 & 0.339 & 0.466 \\ 
        800 & \texttt{T.O.} & 6.381 & 0.573 & 0.555 \\ 
        900 & \texttt{T.O.} & 4.596 & 0.437 & 0.655 \\ 
        1000 & \texttt{T.O.} & 2.427 & 0.591 & 0.768 \\ 
        2000 & \texttt{T.O.} & 17.968 & 1.694 & 1.189 \\ 
        3000 & \texttt{T.O.} & 64.020 & 145.423 & 2.545 \\ 
        4000 & \texttt{T.O.} & 123.153 & 9.462 & 4.451 \\ 
        5000 & \texttt{T.O.} & 55.265 & \texttt{T.O.} & 6.917 \\ 
        6000 & \texttt{T.O.} & 112.407 & 20.635 & 9.897 \\ 
        7000 & \texttt{T.O.} & 90.590 & 29.209 & 12.590 \\ 
        8000 & \texttt{T.O.} & 322.107 & 36.243 & 16.525 \\ 
        9000 & \texttt{T.O.} & 261.887 & \texttt{T.O.} & 22.064 \\ 
        10000 & \texttt{T.O.} & \texttt{T.O.} & 56.249 & 25.341 \\ 
        20000 & \texttt{T.O.} & \texttt{T.O.} & 222.037 & 108.587 \\ 
        30000 & \texttt{T.O.} & \texttt{T.O.} & 498.067 & 247.307 \\ 
        40000 & \texttt{T.O.} & \texttt{T.O.} & \texttt{T.O.} & 448.157 \\ \bottomrule
    \end{tabular}
\end{minipage}
\hfill
\begin{minipage}[t]{0.5\textwidth}
\vspace{0pt}
    \centering
  \newcommand{\opacity}{0.3}
 \begin{tikzpicture}
    \begin{axis}[
    width=\linewidth,
    xmode=log,
    ymode=log,
    xmin=75,
    xmax=45000,
    ymin=0.01,
    ymax=1000,
                   xlabel={Number of data points},
                ylabel={Time (s)},
                label style={font=\small},
                tick label style={font=\scriptsize},
                grid=major,
                legend pos=south east,
                legend style={nodes={scale=0.6, transform shape}},
                log ticks with fixed point,
                xtick={100,500,1000,5000,10000,20000,40000,45000},
                xticklabels={100,500,1000,5000,10000,20000,40000},
                    xticklabel style={
                rotate=45,
               anchor=east,
               },
               ytick={0.1,1,10,100, 600},
                yticklabels={0.1,1,10,100, 600},
             legend columns=2,
            legend style={draw=none, /tikz/every even column/.append style={column sep=0.4cm},
            yshift=-25mm,
            },
    ]

           \addplot[color=blue, mark=*,    opacity=\opacity, mark size=\marksize pt] coordinates {
                (100,1.204) (200,33.720) (300,70.563) (400,16.245) 
                (500,0.631) (600,585.890) (700,250.567)
            };
            \addlegendentry{n.a.}

            \addplot[color=red, mark=square*,    opacity=\opacity, mark size=\marksize pt] coordinates {
                (100,0.115) (200,0.542) (300,0.208) (400,1.126) 
                (500,2.637) (600,3.529) (700,1.055) (800,6.381) 
                (900,4.596) (1000,2.427) (2000,17.968) (3000,64.020) 
                (4000,123.153) (5000,55.265) (6000,112.407) (7000,90.590) 
                (8000,322.107) (9000,261.887) 
            };
            \addlegendentry{n.a., optimized}

    \addlegendimage{color=cyan, mark size=\marksize pt, mark=square*};
\addlegendentry{c.s.};
    \addlegendimage{color=orange, mark size=\marksize pt, mark=triangle*, opacity=\opacity};
\addlegendentry{c.s., optimized};

    \addlegendimage{color=black, only marks, mark size=\marksize * 1.5 pt, mark=x};
\addlegendentry{Timeout};

\addlegendimage{
        color=black,
        mark =o,
        only marks,
        mark options={
        line width=0.45pt,
},
mark size=\marksize pt
};
\addlegendentry{Unknown}

             \addplot[
color=blue,
    mark=x,
     opacity=\opacity,
    only marks, mark size=\marksize*1.5 pt
] coordinates {
    (800,600) (900,600) (1000, 600)
(2000, 600)
(3000, 600)
(4000, 600)
(5000, 600)
(6000, 600)
(7000, 600)
(8000, 600)
(9000, 600)
(10000, 600)
(20000, 600)
(30000, 600)
(40000, 600)
};
             \addplot[
color=red,
    mark=x,
     opacity=\opacity,
    only marks, mark size=\marksize*1.5 pt
] coordinates {
(10000,800)
(20000,800) (30000,800) (40000,800)
};

    \addplot[color=cyan, mark=square*, mark size=\marksize pt] coordinates {
       (200,0.110)
        (300,0.159)
        (400,0.190)
        (500,0.270)
        (600,0.277)
        (700,0.339)
        (800,0.573)
        (900,0.437)
        (1000,0.591)
        (2000,1.694)
        (3000,145.423)
        (4000,9.462)
    };
    \addplot[color=cyan, mark=square*, mark size=\marksize pt] coordinates {
     (6000,20.635)
        (7000,29.209)
        (8000,36.243)
    };

        \addplot[color=cyan, mark=square*, mark size=\marksize pt] coordinates {
        (10000,56.249)
        (20000,222.037)
        (30000,498.067)
    };


    \addplot[color=orange, mark=triangle*, opacity=\opacity, mark size=\marksize pt] coordinates {
        (100,0.045)
        (200,0.054)
        (300,0.252)
        (400,0.101)
        (500,0.128)
        (600,0.163)
        (700,0.466)
        (800,0.555)
        (900,0.655)
        (1000,0.768)
        (2000,1.189)
        (3000,2.545)
        (4000,4.451)
        (5000,6.917)
        (6000,9.897)
        (7000,12.590)
        (8000,16.525)
        (9000,22.064)
        (10000,25.341)
        (20000,108.587)
        (30000,247.307)
        (40000,448.157)
    };
    
    \addplot[
        color=cyan,
        mark =o,
        mark options={
        line width=0.55pt,
    },
        only marks,
        mark size=\marksize pt,
        line width=1pt
    ] coordinates {
        (100,0.02)
    };

    \addplot[
        color=cyan,
        mark=x,
        only marks,
        mark size=\marksize * 1.5 pt] coordinates {
        (5000,700)
        (9000,700)
        (40000,700)
    };

\addplot[gray, domain=\pgfkeysvalueof{/pgfplots/xmin}:
                               \pgfkeysvalueof{/pgfplots/xmax}]
    {600};

    \end{axis}

\end{tikzpicture}
\end{minipage}
\caption{Well-distributedness on the Bank Marketing data in all combinations of non-optimized vs. optimized (opt) baseline, and nested array (n.a.) vs. column-slice (c.s.) data representation. \texttt{T.O.} indicates a timeout (run has exceeded 10 minute runtime). \texttt{UK} indicates that the solver returned \texttt{unknown}. The plot is displayed in log-log-scale.}
\label{ap:well-distributedness}
\end{figure}

\begin{figure}
\centering
\begin{minipage}[t]{0.35\textwidth}
\vspace{0pt}
    \centering
    \begin{tabular}{cccc}
    \toprule
        m & c.s. & n.c.s. & n.a. \\ \midrule
        100 & 0.038 & 0.032 & 0.091 \\ 
        200 & 0.029 & 0.029 & 0.146 \\ 
        300 & 0.029 & 0.036 & 0.218 \\ 
        400 & 0.034 & 0.038 & 0.287 \\ 
        500 & 0.036 & 0.045 & 0.359 \\ 
        600 & 0.042 & 0.048 & 0.451 \\ 
        700 & 0.049 & 0.052 & 0.533 \\ 
        800 & 0.054 & 0.057 & 0.611 \\ 
        900 & 0.055 & 0.060 & 0.696 \\ 
        1000 & 0.062 & 0.060 & 0.791 \\ 
        2000 & 0.109 & 0.108 & 1.718 \\ 
        3000 & 0.154 & 0.160 & 2.721 \\ 
        4000 & 0.220 & 0.227 & 3.718 \\ 
        5000 & 0.298 & 0.307 & 4.894 \\ 
        10000 & 0.892 & 0.902 & 11.280 \\ 
        20000 & 3.105 & 3.132 & 30.040 \\ 
        30000 & 6.866 & 6.982 & 108.980 \\ 
        40000 & 12.972 & 13.307 & 256.190 \\ \bottomrule
    \end{tabular}
\end{minipage}
\hfill
\begin{minipage}[t]{0.5\textwidth}
\vspace{0pt}
    \centering
  \begin{tikzpicture}
     \begin{axis}[
        group style={
        group size=2 by 1,
        horizontal sep=1cm,
    },
    width=\linewidth,
    xmode=log,
    ymode=log,
    xmin=75,
    xmax=45000,
    ymin=0.01,
    ymax=1000,
xlabel={Number of data points},
                ylabel={Time (s)},
                label style={font=\small},
                tick label style={font=\scriptsize},
                grid=major,
                legend pos=south east,
                legend style={nodes={scale=0.6, transform shape}},
                log ticks with fixed point,
                xtick={100,500,1000,5000,10000,20000,40000,45000},
                xticklabels={100,500,1000,5000,10000,20000,40000},
                    xticklabel style={
                rotate=45,
               anchor=east,
               },
               ytick={0.1,1,10,100, 600},
                yticklabels={0.1,1,10,100, 600},
             legend columns=2,
            legend style={draw=none, /tikz/every even column/.append style={column sep=0.8cm},yshift=-20mm,},
    ]

    \addplot[color=blue, mark=square*, mark size=\marksize pt] coordinates {
        (100,0.038)
        (200,0.029)
        (300,0.029)
        (400,0.034)
        (500,0.036)
        (600,0.042)
        (700,0.049)
        (800,0.054)
        (900,0.055)
        (1000,0.062)
        (2000,0.109)
        (3000,0.154)
        (4000,0.220)
        (5000,0.298)
        (10000,0.892)
        (20000,3.105)
        (30000,6.866)
        (40000,12.972)
    };
    \addlegendentry{Column Slice}

    \addplot[color=orange, mark=triangle*, mark size=\marksize pt] coordinates {
        (100,0.032)
        (200,0.029)
        (300,0.036)
        (400,0.038)
        (500,0.045)
        (600,0.048)
        (700,0.052)
        (800,0.057)
        (900,0.060)
        (1000,0.060)
        (2000,0.108)
        (3000,0.160)
        (4000,0.227)
        (5000,0.307)
        (10000,0.902)
        (20000,3.132)
        (30000,6.982)
        (40000,13.307)
    };
    \addlegendentry{Nested Column Slice}

                \addplot[color=green!70!black, mark=triangle*, mark size=\marksize pt] coordinates {
                (100,0.091) (200,0.146) (300,0.218) (400,0.287) (500,0.359)
                (600,0.451) (700,0.533) (800,0.611) (900,0.696) (1000,0.791)
                (2000,1.718) (3000,2.721) (4000,3.718) (5000,4.894) (10000,11.280)
                (20000,30.040) (30000,108.980) (40000,256.190)
            };
    \addlegendentry{Nested Array}
\addplot[gray, domain=\pgfkeysvalueof{/pgfplots/xmin}:
                               \pgfkeysvalueof{/pgfplots/xmax}]
    {600};
    \end{axis}

\end{tikzpicture}
\end{minipage}
\caption{Dataset encodings: nested array (n.a.), nested column-slice (n.c.s.) and column-slice (c.s.) representations for range validity on the Bank Marketing data. The plot is displayed in log-scale.}
\label{ap:dataset-encodings}
\end{figure}
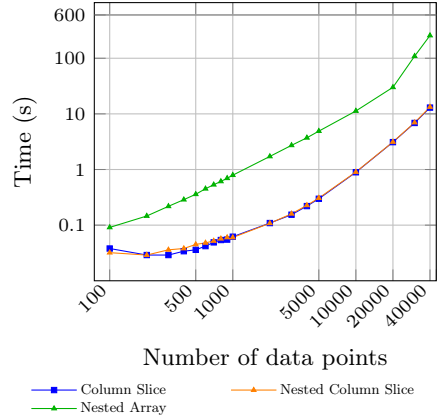

\begin{table}
    \centering
    \caption{Tabular data of the experiment of min-max normalization on the Bank Marketing data belonging to the plot in \Cref{fig:local-properties}.}
    \label{ap:table-min-max}
      \begin{tabular}{ccccc}
    \toprule
        m & Baseline & Recursive & Grounding \\ \midrule
        100 & 0.040 & 0.027 & 0.023 \\ 
        200 & 0.056 & 0.038 & 0.031 \\ 
        300 & 0.082 & 0.044 & 0.034 \\ 
        400 & 0.109 & 0.054 & 0.038 \\ 
        500 & 0.146 & 0.065 & 0.042 \\ 
        600 & 0.180 & 0.072 & 0.048 \\ 
        700 & 0.228 & 0.083 & 0.049 \\ 
        800 & 0.274 & 0.099 & 0.054 \\ 
        900 & 0.331 & 0.107 & 0.060 \\ 
        1000 & 0.388 & 0.119 & 0.062 \\ 
        2000 & 1.279 & 0.321 & 0.106 \\ 
        3000 & 2.718 & 0.646 & 0.158 \\ 
        4000 & 4.648 & 1.105 & 0.220 \\ 
        5000 & 7.031 & 1.677 & 0.299 \\ 
        10000 & 28.784 & 6.205 & 0.888 \\ 
        20000 & 114.133 & 23.499 & 3.093 \\ 
        30000 & 263.310 & 51.932 & 6.994 \\ 
        40000 & 500.127 & 93.410 & 13.160 \\ \bottomrule
\end{tabular}
\end{table}

\begin{table}
    \centering
        \caption{Tabular data of the experiment of class balance on the Bank Marketing data belonging to the plot in \Cref{fig:local-properties}. \texttt{T.O.} indicates a timeout (run has exceeded 10 minute runtime).}
    \label{ap:table-class-balance}
    \begin{tabular}{ccccc}
    \toprule
        m & Baseline & Grounding \\ \midrule
        100 & 0.210 & 0.022 \\ 
        200 & 0.564 & 0.019 \\ 
        300 & 1.212 & 0.019 \\ 
        400 & 2.033 & 0.023 \\ 
        500 & 3.128 & 0.022 \\ 
        600 & 4.511 & 0.025 \\ 
        700 & 6.047 & 0.026 \\ 
        800 & 7.835 & 0.027 \\ 
        900 & 10.133 & 0.028 \\ 
        1000 & 12.671 & 0.031 \\ 
        2000 & 58.257 & 0.042 \\ 
        3000 & 149.227 & 0.052 \\ 
        4000 & 303.570 & 0.064 \\ 
        5000 & 521.943 & 0.071 \\ 
        10000 & \texttt{T.O.} & 0.120 \\ 
        20000 & \texttt{T.O.} & 0.207 \\ 
        30000 & \texttt{T.O.} & 0.283 \\ 
        40000 & \texttt{T.O.} & 0.393 \\ \bottomrule
    \end{tabular}
\end{table}

\begin{table}
    \centering
    \caption{Tabular data of the experiment of range validity on the Bank Marketing data belonging to the plot in \Cref{fig:global-properties}. For all properties (except Grounding (c.s.)), the data was encoded as a nested array. In Grounding (c.s.), it is encoded in the column-slice representation. \texttt{T.O.} indicates a timeout (run has exceeded 10 minute runtime).}
    \label{ap:table-range-validity}
        \begin{tabular}{cccccc}
    \toprule
        m & Baseline  & Recursive & Grounding & Grounding (c.s.) \\ \midrule
        100 & 0.139 & 0.101 & 0.091 & 0.038 \\ 
        200 & 0.315 & 0.149 & 0.146 & 0.029 \\ 
        300 & 0.591 & 0.223 & 0.218 & 0.029 \\ 
        400 & 0.916 & 0.296 & 0.287 & 0.034 \\ 
        500 & 1.300 & 0.370 & 0.359 & 0.036 \\ 
        600 & 1.854 & 0.469 & 0.451 & 0.042 \\ 
        700 & 2.402 & 0.560 & 0.533 & 0.049 \\ 
        800 & 2.985 & 0.644 & 0.611 & 0.054 \\ 
        900 & 3.707 & 0.742 & 0.696 & 0.055 \\ 
        1000 & 4.422 & 0.839 & 0.791 & 0.062 \\ 
        2000 & 15.829 & 1.923 & 1.718 & 0.109 \\ 
        3000 & 35.152 & 3.187 & 2.721 & 0.154 \\ 
        4000 & 59.117 & 4.534 & 3.718 & 0.220 \\ 
        5000 & 99.707 & 6.203 & 4.894 & 0.298 \\ 
        10000 & 458.763 & 15.985 & 11.280 & 0.892 \\ 
        20000 & \texttt{T.O.} & 51.174 & 30.040 & 3.105 \\ 
        30000 & \texttt{T.O.} & 165.577 & 108.980 & 6.866 \\ 
        40000 & \texttt{T.O.} & 372.547 & 256.190 & 12.972 \\ \bottomrule
    \end{tabular}
\end{table}


\end{document}